\documentclass[aps, prl, twocolumn, titlepage, showpacs]{revtex4}

\usepackage{graphicx}
\usepackage{color}
\usepackage{dcolumn}
\usepackage{amsfonts}
\usepackage{bm}

\usepackage{amsmath}
\usepackage{delarray}

\begin{document}
\title{Phonon spectra of a two-dimensional solid dusty plasma modified by two-dimensional periodic substrates}

\author{Y. Huang$^1$, W. Li$^{2, 3}$, C. Reichhardt$^4$, C. J. O. Reichhardt$^4$, and Yan Feng$^{1, \ast}$}
\affiliation{
$^1$ Center for Soft Condensed Matter Physics and Interdisciplinary Research, School of Physical Science and Technology, Soochow University, Suzhou 215006, China\\
$^2$ School of Science, Nantong University, Nantong 226019, China\\
$^3$ Jiangsu Key Laboratory of Thin Films, Soochow University, Suzhou 215006, China\\
$^4$ Theoretical Division, Los Alamos National Laboratory, Los Alamos, New Mexico 87545, USA\\
$\ast$ E-mail: fengyan@suda.edu.cn}

\date{\today}

\begin{abstract}
Phonon spectra of a two-dimensional (2D) solid dusty plasma modulated by 2D square and triangular periodic substrates are investigated using Langevin dynamical simulations.
The commensurability ratio, i.e., the ratio of the number of
particles to the number of potential well minima, is set to 1 or 2.
The resulting phonon spectra show that
propagation of
waves is always suppressed
due to the confinement of particles by the applied 2D periodic substrates.
For
a commensurability ratio of 1, the
spectra indicate that all particles mainly oscillate at one specific frequency, corresponding to the harmonic oscillation frequency of one single particle inside one potential well. 
At a commensurability ratio of 2, the substrate allows two particles
to sit
inside
the bottom of each potential well, and the resulting longitudinal and transverse spectra exhibit four branches in total.
We find that the two moderate branches come from the harmonic oscillations of one single particle and two combined particles in the potential well.
The other two branches correspond to the relative motion of the two-body structure in each potential well in the radial and azimuthal directions.
The difference in the spectra between the square and triangular substrates is attributed to the anisotropy of the substrates and the resulting alignment directions of the two-body structure in each potential well.

\end{abstract}

\maketitle
\section{\uppercase\expandafter{\romannumeral1}. Introduction}
Dynamical behaviors of collective particles modulated by substrates are of great interest and
have been widely studied in various two-dimensional (2D) systems, such as colloidal monolayers~\cite{Reichhardt:2005}, vortices in type-\uppercase\expandafter{\romannumeral2} superconductors~\cite{harada:1996}, electron crystals on a liquid helium surface~\cite{monceau:2012}, pattern-forming systems~\cite{Reichardt:2003, sengupta:2003}, and dusty plasmas~\cite{gu:2020}.
When a substrate is applied to these systems, a variety of new physical phenomena can be generated, such as
pinning and depinning dynamics~\cite{Rechardt:2017},
Shapiro steps~\cite{tekic:2010},
phase transitions~\cite{mandelli:2015}, and
anomalous transport~\cite{shaina:2017}.
These external substrates include one-dimensional (1D) periodic substrates~\cite{Reichardt:2015}, 2D periodic substrates~\cite{bechinger:2001}, quasiperiodic substrates~\cite{bohlein:2012}, quasicrystalline substrates~\cite{su:2017}, and random substrates~\cite{Pertsinidis:2008}.
More interesting phenomena are currently being explored in a range of various systems
for these substrates.

A dusty plasma~\cite{thomas:1996, juan:1996, melzer:1996, fortov:2005, piel:2010, bonitz:2010, merlino:2004}, also called a complex plasma, is a mixture of free electrons, ions, neutral gas atoms and micron-sized dust particles. 
Under typical laboratory conditions, by absorbing free electrons and ions in plasmas, micron-sized dust particles gain a high negative charge of $\approx 10^{-4} e$ in the steady state within microseconds.
Due to their high negative charge, these dust particles are strongly coupled,
and can be self-organized into a single layer~\cite{feng:2011, qiao:2014}, i.e., the 2D dusty plasma, exhibiting typical solid-like~\cite{feng:2008, hartmann:2014} or liquid-like~\cite{thomas:2004, feng:2010} properties.
The interparticle interaction between these dust particles can be described as
a Yukawa repulsion~\cite{kono:2000}, where the shielding effect comes from the free electrons and ions.
As a promising physical model system
permitting the direct imaging of individual dust particles, various fundamental physical processes of solids and liquids, such as diffusion~\cite{don:2009, liu:2008}, shear viscosity~\cite{nosenko:2004},
and phase transitions~\cite{feng:2008} have been studied widely at the kinetic level in dusty plasmas.

Phonon spectra are often calculated in the investigations of dusty plasmas
using the velocities and positions of dust particles from either experimental observations~\cite{nunomura:2002, nosenko:2006, nosenko:2003, liu:2009, hartmann:2009} or computer simulations~\cite{hou:2009, goree:2012}. 
These spectra provide the energy distribution of phonons
in ${\bf k} - \omega $ space, corresponding to the dispersion relation~\cite{nunomura:2002} of the studied system.
In dusty plasmas, the phonon spectra can be derived directly from the thermal motion of the dust particles~\cite{nunomura:2002},
in good agreement with the theoretical dispersion relations~\cite{wang:2001}.
In addition to the
2D dusty plasmas~\cite{nunomura:2002, nosenko:2006, nosenko:2003, liu:2009, hartmann:2009}, phonon spectra
have also been studied for a 1D chain~\cite{Liu:2003} and
a ring~\cite{Sheridan:2016} of dusty plasmas.

Recently, 1D periodic substrates have been introduced in dusty plasmas to modulate the collective behaviors of dust particles
in Langevin dynamical simulations. 
In Ref.~\cite{wang:2018}, as the 1D periodic substrate depth increases gradually from zero, it is found that the 2D dusty plasma exhibits
structural transitions
from a disordered liquid state to a modulated ordered state, and finally to a modulated disordered state.
As the width of the 1D periodic substrate
is gradually varied, the particle diffusion exhibits
an oscillation-like feature~\cite{Li:2020}.
When a gradually increasing external driving force is applied to the
2D dusty plasma on the 1D periodic substrate,
three different states, i.e., pinned, disordered plastic flow,
and moving ordered states,
appear~\cite{Li:2019}.
The properties of the transition between these states are determined by the depth of the 1D periodic substrate~\cite{gu:2020}.
In addition, the phonon spectra of a 2D dusty plasma modulated by a 1D periodic substrate are studied in~\cite{Li:2018}, where breathing spectra and the backward propagation of sloshing spectra are observed.
However, the collective dynamics of a 2D dusty plasma modulated by 2D periodic substrates,
and the corresponding phonon spectra, have not been studied
previously.

Another system of particles interacting with 2D periodic substrates that has been studied extensively is charged colloids coupled to optical or patterned square or triangular arrays, where various commensuration effects appear when the number of particles is an integer multiple of the number of substrate minima~\cite{Reichhardt:2002, brunner:2002, agra:2004, sarlah:2005, reichhardt:2009, shawish:2011, chern:2013, neuhaus:2013, ambriz:2016, gallo:2021}. One of the goals in studying these systems is to create structures with phononic band gap properties, similar to photonic band gaps, in which certain mechanical waves cannot propagate through the system while others can~\cite{kushwaha:1993, croenne:2011}. In colloidal assemblies, investigations have considered how a periodic substrate could be used to create such phononic band gaps~\cite{baumgartl:2007}; however, the phononic modes in most colloidal systems are strongly damped, giving only a very limited range of propagation~\cite{baumgartl:2008}. In contrast, the reduced damping in dusty plasmas can produce much stronger phononic modes, so understanding how a periodic substrate could create phononic band gaps in dusty plasmas could also provide insight into how such phononic band gaps would appear in other systems. Examples of underdamped systems include charged colloidal particles suspended in air rather than in a solution and interacting with an array of optical traps~\cite{hwang:2020}, ions in periodic atom traps~\cite{bruzewicz:2016}, or even a Wigner crystal in a monolayer system~\cite{smolenski:2021}. Our work indicates that a 2D periodic substrate can create phononic band gaps, and our results could be general to a wide range of underdamped systems coupled to a 2D periodic substrate.

This
paper is organized as follows. 
In Sec.~\uppercase\expandafter{\romannumeral2}, we briefly describe our Langevin simulation method to mimic solid 2D dusty plasmas under 2D periodic square and triangular substrates.
In Sec.~\uppercase\expandafter{\romannumeral3}, we present the phonon spectra of the 2D dusty plasma on different types of 2D periodic substrates. 
We find that the phonon spectra of
the 2D solid dusty plasma
changes to branches with nearly
unmodified
frequency values, suggesting that all particles are mainly confined by the substrate. The frequencies of these branches
agree well with our derivation of the oscillation modes of
dust particles within potential wells of the 2D substrates.
Finally, we summarize our findings in Sec.~\uppercase\expandafter{\romannumeral4}.

\section{\uppercase\expandafter{\romannumeral2}. Simulation method}
Traditionally, dusty plasma can be characterized by two dimensionless parameters~\cite{San:2001,Oh:2000}, the coupling parameter $\Gamma= Q^2/(4 \pi \epsilon_0 a k_B T)$ and the screening parameter $\kappa= a / \lambda_D$. Here $Q$ is the charge of one particle, $T$ is the averaged kinetic temperature of the particles, $a=(\pi n)^{-1/2}$ is the Wigner-Seitz radius~\cite{kal:2004} with the 2D areal number $n$, and $\lambda_D$ is Debye screening length. The Wigner-Seitz radius $a$, and the lattice constant $b$, i.e., the average distance between nearest neighbors ($b = 1.9046a$ for the 2D triangular lattice), are both used to normalize the length.

We use Langevin dynamical simulations to investigate the dynamics of a 2D dusty plasma on 2D periodic substrates. In our simulations, for each particle $i$, the equation of motion~\cite{Li:2018} is
\begin{equation}\label{equal_1}
{	m \ddot{\bf r}_i = -\nabla \Sigma \phi_{ij} - \nu m\dot{\bf r}_i + \xi_i(t)+{\bf F}^{S}_i.}
\end{equation}
The first term on the right-hand side of Eq.~(\ref{equal_1}) comes from the binary Yukawa repulsion~\cite{Liu:2003}, $\phi_{ij}=Q^{2} \exp (-r_{i j} / \lambda_{D})/ 4 \pi \epsilon_{0} r_{i j}$, where $r_{ij}$ is the distance between two dust particles $i$ and $j$.
The second and third terms correspond to the frictional drag $- \nu m\dot{\bf r}_i$ and the Langevin random kicks $\xi_i(t)$~\cite{Gun:1982, Feng:2008}, respectively. The last term is the force from the applied 2D substrate, as we explain in detail later.

Our simulation parameters are listed below. We simulate $N_p = 1024$ particles, confined in a $61.1a \times 52.9a$ rectangular box with
periodic boundary conditions. The conditions of the 2D dusty plasma are specified as  $\Gamma = 1000$ and $\kappa = 2$, corresponding to the typical solid state of 2D Yukawa systems~\cite{ha:2005}. The frictional drag coefficient is specified as $\nu / {\omega}_{pd} = 0.027$, close to the typical experimental value~\cite{feng:2011}, where ${\omega}_{pd} = (Q^2/2\pi\varepsilon_0 m a^3)^{1/2}$ is the nominal dusty plasma frequency~\cite{kal:2004}. For each simulation run, we integrate $\geq 10^7$ steps with the time step of $0.003{\omega}_{pd}^{-1}$  to obtain the positions and velocities of all particles.

\begin{figure}[htb]
    \centering
    \includegraphics{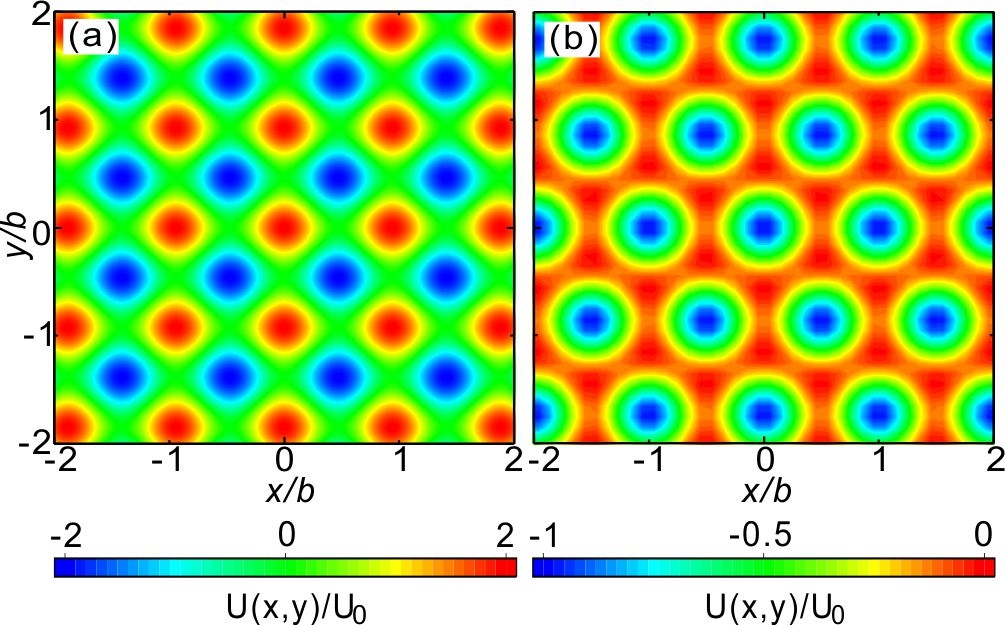}
    \caption{\label{fig:1}
   Contour plot of the applied potential for the square (a) and triangular (b) substrates in our simulations. Here, only $\approx 1.8\%$ of the total simulation box is shown.
    }
\end{figure}

We investigate the effects
on the dynamics of 2D dusty plasmas
of two types
of substrates,
square and triangular,
as shown in Fig.~\ref{fig:1}.
The square substrate~\cite{Reichhardt:2002}
has the form
\begin{equation}\label{equal_2}
{ U(x, y) = U_0 [\cos(2\pi x/w)+\cos(2\pi y/w)],}
\end{equation}
where $U_0$ and $w$ correspond to the depth and width of the potential wells, in units of $E_0=Q^{2}/4 \pi \epsilon_{0} b $ and $b$, respectively.
The triangular substrate~\cite{Davi:2017} is
given by 
\begin{equation}\label{equal_3}
\begin{aligned}
 U(x, y) = &-\frac{2}{9} U_0 [\frac{3}{2}+2 \cos(\frac{2\pi x}{w}) \cos(\frac{2\pi y}{\sqrt3 w})\\
&+\cos(\frac{4\pi y}{\sqrt3 w})],
\end{aligned}
\end{equation}
where $U_0$ and $w$ are related to the depth of the potential wells and the distance between them.
From these substrate definitions, we can easily derive the forces ${\bf F}^{S}_i$ acting on the dust particle $i$ as
\begin{equation}
\label{equal_4}
\begin{aligned}
{\bf F}^{S}_i = \frac {2\pi U_0} {w} \sin (\frac{2\pi x}{w}) \hat{\bf x} + \frac {2\pi U_0} {w} \sin (\frac{2\pi y}{w}) \hat{\bf y}
\end{aligned}
\end{equation}
from the square substrate, and 
\begin{equation}
\label{equal_5}
\begin{aligned}
{\bf F}^{S}_i = &-\frac {8\pi U_0} {9w} \sin (\frac{2\pi x}{w}) \cos (\frac{2\pi y}{\sqrt3 w}) \hat{\bf x} \\
&- \frac {8\pi U_0} {9\sqrt3 w} [\cos (\frac{2\pi x}{w}) \sin (\frac{2\pi y}{\sqrt3 w}) + \sin (\frac{4\pi y}{\sqrt3 w})] \hat{\bf y} 
\end{aligned}
\end{equation}
from the triangular substrate, respectively.

In our simulations, we choose the commensurability ratio $\rho$
defined as $\rho = N_p/N_w$
to be either 1 or 2.
Here, $N_p$ is the total
number of particles
while $N_w$ is the number of potential well minima. Since we simulate 1024 particles,
we need to arrange either 1024 or 512 potential wells
in our simulation box.
For the square substrate, the substrate parameter $w$ is specified as $\approx 0.93 b$ and $\approx 1.32 b$, corresponding to $\rho = 1$ and $2$, respectively. For the triangular substrate, the substrate parameter $w$ is specified as $b$ and $1.39b$, corresponding to $\rho = 1$ and $2$, respectively. We specify the other parameter of the substrate as $U_0 = 0.5E_0$ and $E_0$, respectively. We note that since our simulation box is designed to match the length ratio of the triangular lattice and not that of the square lattice,
the parameter $w$ in the $x$ and $y$ directions of Eq.~(\ref{equal_2}) varies slightly, $\approx 1\%$, for the square substrate
in order to satisfy the periodic boundary conditions,
while the triangular substrate does not have this problem.

To obtain the phonon spectra, we use the Fourier transforms of the longitudinal and transverse current correlation function. 
The current autocorrelation functions are defined as~\cite{oh:2000_1,Liuh:2003}
\begin{equation}\label{equal_6}
{	C_{L}({\bf k},t)= \frac {1}{N_p} \langle [ {\bf k} \cdot {\bf j}({\bf k},t)][{\bf k} \cdot {\bf j}(-{\bf k},0)] \rangle ,}
\end{equation}
for the longitudinal mode, and 
\begin{equation}\label{equal_7}
{	C_{T}({\bf k},t)= \frac {1}{2N_p} \langle [ {\bf k} \times {\bf j}({\bf k},t)]\cdot [{\bf k} \times {\bf j}(-{\bf k},0)] \rangle ,}
\end{equation}
for the transverse mode. Here, $\bf k$ is the wave vector, and ${\bf j}({\bf k},t) = \sum_{j=1}^{N_p} {\bf v}_{j}(t) \exp [i{\bf k}\cdot {\bf r}_j(t)]$ is the current function for a given wave vector $\bf k$, where ${\bf v}_{j}(t)$ and ${\bf r}_j(t)$ are the velocity and position of the $j$th particle, respectively. Finally, the phonon spectra can be obtained by the Fourier transform of these current autocorrelation functions (\ref{equal_6}) and (\ref{equal_7}), defined as
\begin{equation}\label{equal_8}
{	\tilde{C}_{L,T}({\bf k},\omega) = \int_0^\infty e^{-i\omega t}C_{L,T}({\bf k},t)\mathrm{d}t}.
\end{equation}
Here, $\tilde{C}_{L}({\bf k},\omega)$ and $\tilde{C}_{T}({\bf k},\omega)$ are the longitudinal and transverse wave spectra, respectively.
Due to the anisotropy
which appears in the 2D Yukawa solids when they are modified
by the 2D substrate,
we need to analyze the phonon spectra in different directions. Here, we focus on the phonon spectra of $\tilde{C}_{L}({\bf k}_x,\omega)$, $\tilde{C}_{T}({\bf k}_x,\omega)$, $\tilde{C}_{L}({\bf k}_y,\omega)$, and $\tilde{C}_{T}({\bf k}_y,\omega)$, corresponding to the longitudinal and transverse spectra with wave vector along the $x$ and $y$ directions, respectively.

\section{\uppercase\expandafter{\romannumeral3}. Results and discussions}

\subsection{ {\bf A}. Particle arrangement under substrates}

\begin{figure}[htb]
    \centering
    \includegraphics{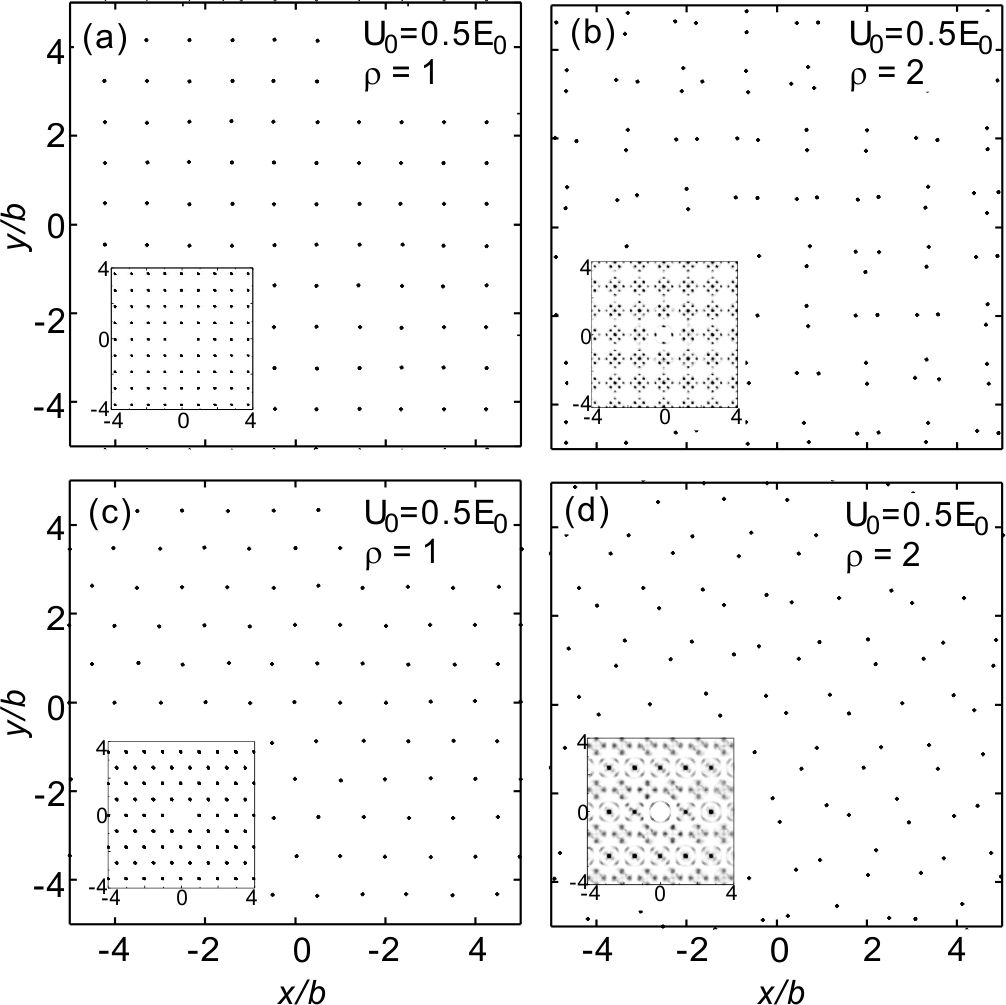}
    \caption{\label{fig:2}
Snapshots of the particle positions from our simulations on the square (a, b) and triangular (c, d) substrates of the same depth, with commensurability ratio $\rho$ values of either 1 or 2. Here, $\rho$ is defined as $\rho = N_p/N_w$, the ratio of the number of particles $N_p$ to the number of potential wells $N_w$. For each panel, the inset in the lower left corner corresponds to the calculated 2D distribution function $g(x,y)$ of the particles.
    }
\end{figure}

In Fig.~\ref{fig:2}, we present
snapshots of particle positions from our simulations showing 
the arrangement of particles under the square and triangular substrates.
For the square substrate, when the commensurability ratio $\rho = 1$ in Fig.~\ref{fig:2}(a), all particles are pinned
at the bottom of potential wells, forming
an ordered square arrangement. When the commensurability ratio $\rho = 2$ in Fig.~\ref{fig:2}(b),
most potential wells contain two particles.
At the bottom of each potential well, the two particles repel each other due to their Yukawa repulsion, forming a typical two-body structure similar to the colloidal molecular crystals or vortex molecular crystals found for colloidal particles \cite{Reichhardt:2002, brunner:2002, agra:2004, sarlah:2005, reichhardt:2009, shawish:2011} or superconducting vortices on 2D periodic substrate arrays \cite{reichhardt:2007, neal:2007}. 
Interestingly, in Fig.~\ref{fig:2}(b), within each potential well, the two particles are mostly aligned in two directions, parallel to either the $x$ or the $y$ directions, corresponding to the two axes of the square substrate. 
The inset of each panel presents the corresponding 2D distribution function~\cite{loudiyi:1992} $g(x,y)$, which provides the probability density of finding a particle at the relative position $(x,y)$ from one chosen particle. 
From our calculated $g(x,y)$ in Figs.~\ref{fig:2}(a) and ~\ref{fig:2}(b), we find that, under the square substrate, the structure along the $x$ direction is nearly the same as that along the $y$ direction, so that the phonon spectra of $\tilde{C}_{L}({\bf k}_x,\omega)$ should be nearly the same as
those for
$\tilde{C}_{L}({\bf k}_y,\omega)$, and similarly for $\tilde{C}_{T}({\bf k}_x,\omega)$ and $\tilde{C}_{T}({\bf k}_y,\omega)$, as we will verify later.
We note that a few potential wells contain either three or one particles,
probably due to an energy fluctuation, as shown in Fig.~\ref{fig:2}(b).

\par
Under the triangular substrate, the arrangement of particles is completely different from
that found for
the square substrate. When the commensurability ratio $\rho = 1$ in Fig.~\ref{fig:2}(c), the particles are pinned
at the bottom of the potential well, forming a triangular lattice with hexagonal symmetry
matching the triangular substrate. When the commensurability ratio $\rho = 2$ in Fig.~\ref{fig:2}(d),
most of the potential wells contain two particles forming a
two-body structure,
similar to Fig.~\ref{fig:2}(b). The alignment direction of
the pairs of particles in each potential well in Fig.~\ref{fig:2}(b) is either roughly
parallel, $60^{\circ}$, or $120^{\circ}$ with respect to the $x$ direction,
corresponding to the principal axes of the triangular substrate.
We note that under thermal motion,
these particle pairs in the triangular substrate are more likely to rotate,
unlike the particle pairs on the square substrate described above.
From $g(x,y)$ in Figs.~\ref{fig:2}(c) and \ref{fig:2}(d), we find the anisotropy of the static structure on the triangular substrate,
which is distinct from
those in Figs.~\ref{fig:2}(a) and \ref{fig:2}(b). As a result, the corresponding longitudinal spectra $\tilde{C}_{L}({\bf k},\omega)$ (transverse spectra $\tilde{C}_{T}({\bf k},\omega)$) along the $x$ direction should be completely different from those along the $y$ direction.
Similar to Fig.~\ref{fig:2}(b), there are also a few potential wells containing either three or only one particle in Fig.~\ref{fig:2}(d).

\subsection{{\bf B}. Wave spectra under a square substrate}

\begin{figure}[htb]
    \centering
    \includegraphics{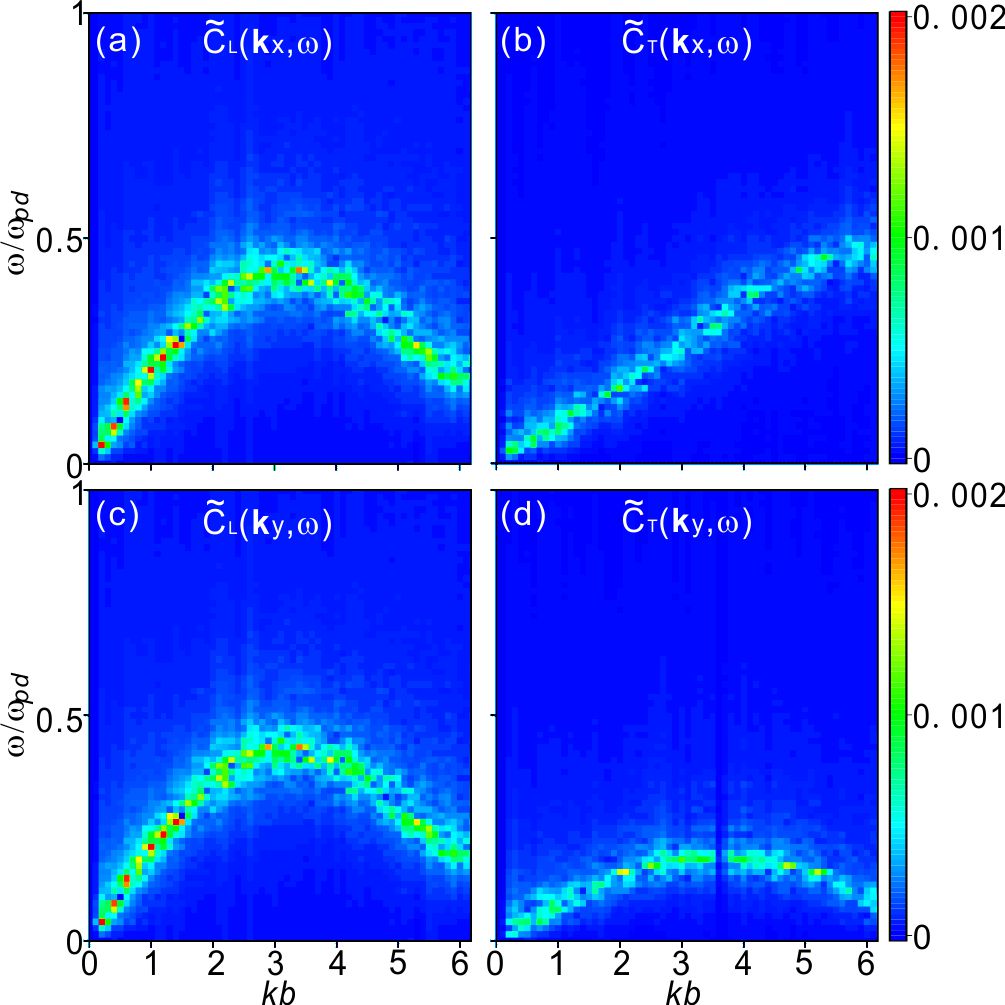}
    \caption{\label{fig:3}
      Calculated longitudinal $\tilde{C}_{L}({\bf k}_x,\omega)$ (a), $\tilde{C}_{L}({\bf k}_y,\omega)$ (c), and transverse $\tilde{C}_{T}({\bf k}_x,\omega)$ (b), $\tilde{C}_{T}({\bf k}_y,\omega)$ (d) phonon spectra for our simulated 2D Yukawa solid without any substrates. The difference between these two transverse spectra
      arises from the anisotropy of the triangular structure of the 2D Yukawa solid,
as does the longitudinal spectra difference. The conditions of our simulated 2D Yukawa solid are $\Gamma = 1000$, $\kappa = 2.0$, and $\nu = 0.027 \omega_{pd}$.
    }
\end{figure}

\begin{figure}[htb]
    \centering
    \includegraphics{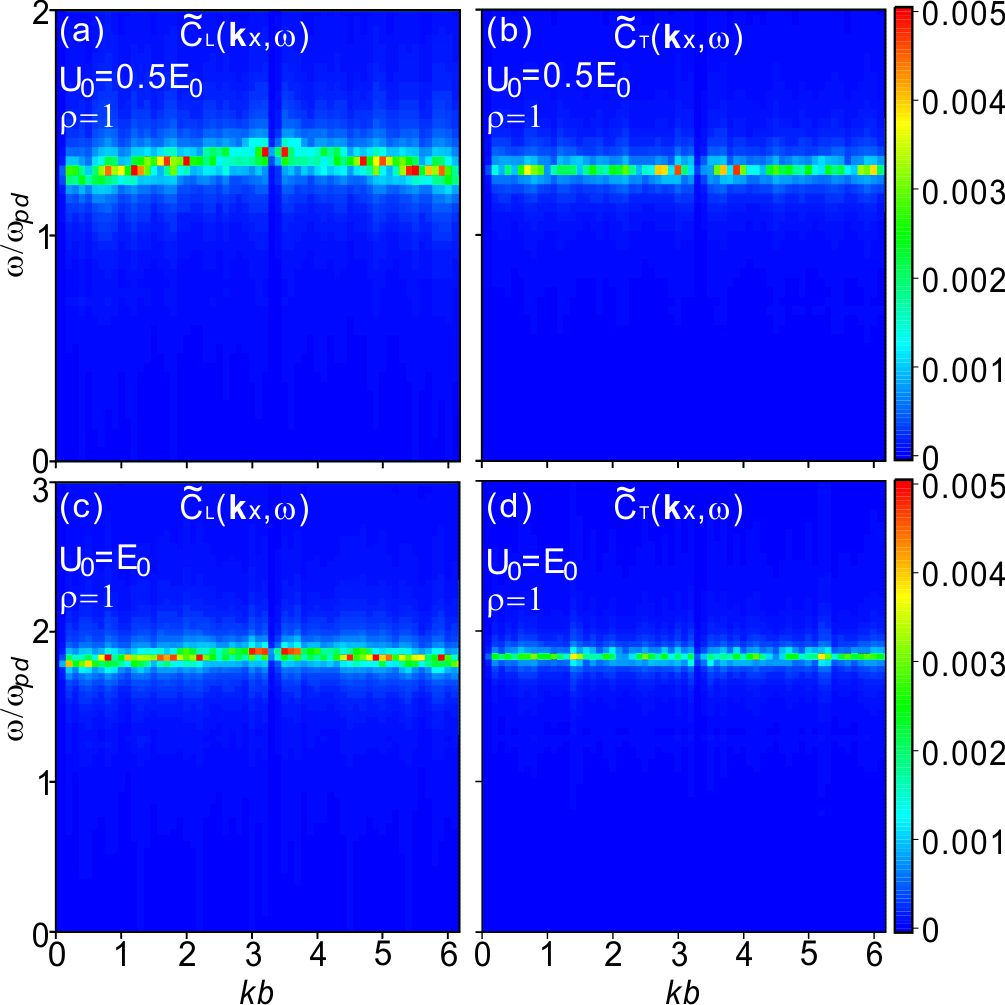}
    \caption{\label{fig:4}
      Calculated longitudinal $\tilde{C}_{L}({\bf k}_x,\omega)$ (a, c) and transverse spectra $\tilde{C}_{T}({\bf k}_x,\omega)$ (b, d) of the 2D Yukawa solid, under the periodic square substrate with different depths of $U_0=0.5 E_0$ in (a, b) and $U_0 = E_0$ in (c, d), with a commensurability ratio of $\rho = 1$. Clearly, for each condition, both the longitudinal and transverse spectra mainly concentrate around one specific frequency, suggesting that the wave propagation is suppressed due to the confinement from the substrate. As the depth of the substrate increases from $U_0=0.5 E_0$ to $E_0$, the frequency of the wave spectra is
enhanced. The two frequency values in these four panels agree well with the harmonic oscillation frequencies estimated from one single particle inside a potential wells with $U_0=0.5 E_0$ and $E_0$.
    }
\end{figure}

\begin{figure}[htb]
    \centering
    \includegraphics{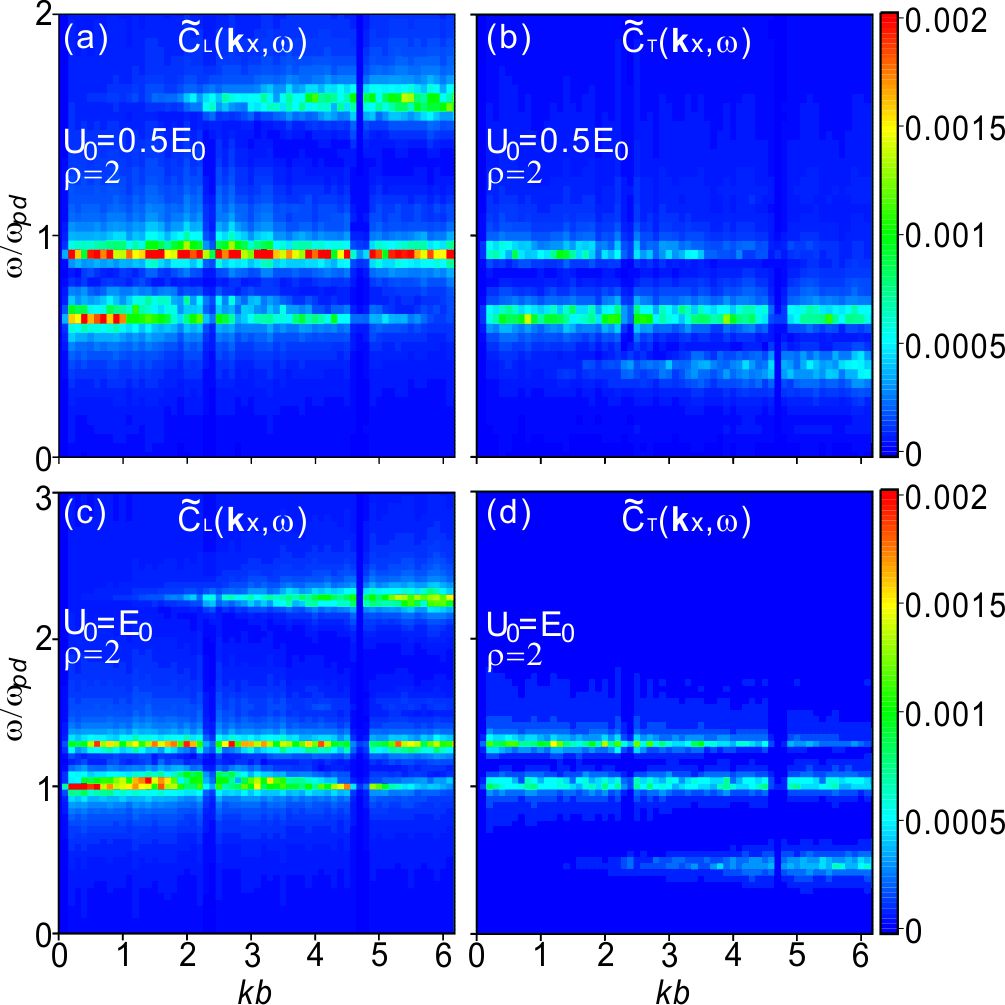}
    \caption{\label{fig:5}
      Calculated longitudinal $\tilde{C}_{L}({\bf k}_x,\omega)$ (a, c) and transverse phonon spectra $\tilde{C}_{T}({\bf k}_x,\omega)$ (b, d) of the 2D Yukawa solid, under periodic square substrates with $U_0=0.5 E_0$ in (a, b) and $U_0 = E_0$ in (c, d), for a commensurability ratio of $\rho = 2$. Clearly, both the longitudinal and transverse spectra mainly concentrate around three frequencies, due to the motion modes of the two particles within each potential well. For the substrate depth $U_0=0.5 E_0$, both the longitudinal (a) and transverse spectra (b) have two
equal frequency values, agreeing well with the oscillation frequencies of one single particle and two combined particles within the potential well of $U_0 = 0.5E_0$. In addition, the highest frequency of the longitudinal spectra (a) and the lowest frequency of the transverse spectra (b) are consistent with the oscillation frequencies of the relative motion of a two-body structure inside the potential well, in the radial and azimuthal directions, respectively. As the depth of the substrate increases to $E_0$, the frequencies of the wave spectra in (c, d) also agree with the estimated frequencies from $U_0 = E_0$. 
    }
\end{figure}

To better quantify the substrate effect,
we first calculate the phonon spectra of a 2D Yukawa solid with the same values of $\Gamma$ and $\kappa$ without any substrates, as presented in Fig.~\ref{fig:3}.
For both the longitudinal spectra $\tilde{C}_{L}({\bf k}_x,\omega)$ in Fig.~\ref{fig:3}(a) and the transverse spectra $\tilde{C}_{T}({\bf k}_x,\omega)$ in Fig.~\ref{fig:3}(b), the slope around the smaller wavenumbers indicates the longitudinal and transverse wave propagation speeds.
The longitudinal spectra of $\tilde{C}_{L}({\bf k}_y,\omega)$ in Fig.~\ref{fig:3}(c) are similar to $\tilde{C}_{L}({\bf k}_x,\omega)$ in Fig.~\ref{fig:3}(a), for almost the full range of the wavenumber. The transverse spectra of $\tilde{C}_{T}({\bf k}_y,\omega)$ in Fig.~\ref{fig:3}(d) are similar to $\tilde{C}_{T}({\bf k}_x,\omega)$ in Fig.~\ref{fig:3}(b) only in the lower wavenumbers. However, when the wavenumber is higher, $\tilde{C}_{T}({\bf k}_y,\omega)$ in Fig.~\ref{fig:3}(d) are quite different from $\tilde{C}_{T}({\bf k}_x,\omega)$ in Fig.~\ref{fig:3}(b)
due to the anisotropy of the highly ordered triangular lattice at lower temperatures.

In Fig.~\ref{fig:4}, we present our calculated phonon spectra of the 2D Yukawa solid under square substrates with $U_0 = 0.5E_0$ and $E_0$, respectively, for a commensurability ratio of $\rho = 1$. In Fig.~\ref{fig:4}, the nearly unchanged frequency for all wavenumbers in the longitudinal and transverse spectra for each $U_0$ shows that the propagation of both longitudinal and transverse waves are strongly suppressed, indicating that the corresponding group velocity is nearly zero, i.e., all particles only oscillate locally with this frequency. When the substrate depth increases from $0.5 E_0$ to $E_0$, the wave propagation is further suppressed, as shown in Fig.~\ref{fig:4}.
In addition, we verify that the phonon spectra of $\tilde{C}_{L}({\bf k}_x,\omega)$ ($\tilde{C}_{T}({\bf k}_x,\omega)$) are almost exactly the same as those of $\tilde{C}_{L}({\bf k}_y,\omega)$ ($\tilde{C}_{T}({\bf k}_y,\omega)$).

For each substrate depth, the frequencies of the longitudinal $\tilde{C}_{L}({\bf k}_x,\omega)$ and transverse spectra $\tilde{C}_{L}({\bf k}_x,\omega)$ are almost the same, since they
correspond to the motion of particles in the $x$ and $y$ directions. 
As the substrate depth increases from $U_0=0.5E_0$ to $E_0$, this frequency is greatly enhanced from $\approx 1.29 \omega_{pd}$ in Figs.~\ref{fig:4}(a) and \ref{fig:4}(b) to $\approx 1.79 \omega_{pd}$ in Figs.~\ref{fig:4}(c) and \ref{fig:4}(d). 
In fact, these two frequencies can be derived from the harmonic oscillation of a single particle within the bottom of the square potential well of Eq.~(\ref{equal_2}) with depth values of $0.5E_0$ and $E_0$, respectively. 
Since the particles only vibrate around the bottom of potential well, we can linearize the force from the square potential well of Eq.~(\ref{equal_4}) to yield the spring constant of $k_s =  4 \pi^{2} U_0/ w^2 $, where the subscript $s$ refers to the square potential well.
Thus, the oscillation frequency of a single particle is just $\omega_{1} =\sqrt {k_s / m}= \sqrt{ 4 \pi^{2} U_0/ m w^2 }$. 
Substituting the depth $U_0$ and width $w$ of the potential well into this oscillation frequency equation, we derive the oscillation frequency of $1.28 \omega_{pd} $ and $1.82 \omega_{pd}$ for the substrate depth of $0.5E_0$ and $E_0$, respectively. 
Clearly, these two derived frequency values of $1.28 \omega_{pd} $ and $1.82 \omega_{pd}$ agree well with the phonon spectra frequencies in Fig.~\ref{fig:4}. 
We note that the slight difference between the derived frequencies and those shown in Fig.~\ref{fig:4} may come from the interparticle interaction, which is not included in the derivation of the single particle harmonic oscillation.

In Fig.~\ref{fig:5}, we present our calculated phonon spectra of the 2D Yukawa solid under the same substrate of depths $U_0 = 0.5E_0$ and $ E_0$ for a commensurability ratio $\rho = 2$.
Compared with the spectra for $\rho = 1$ in Fig.~\ref{fig:4}, all spectra in each panel of Fig.~\ref{fig:5}
contain three branches, corresponding to three different modes of the particle motion, probably due to the two-body structure within each potential well. The frequency of each branch is nearly unchanged while the wavenumber varies, suggesting that the corresponding wave propagation is also strongly suppressed as in Fig.~\ref{fig:4}. Interestingly, for the same substrate depth $U_0$, the frequencies of two branches in $\tilde{C}_{L}({\bf k}_x,\omega)$ and $\tilde{C}_{T}({\bf k}_x,\omega)$ are the same, such as $0.91\omega_{pd} $ and $0.62 \omega_{pd} $ in Figs.~\ref{fig:5}(a, b), as well as $1.25 \omega_{pd}$ and $1.00 \omega_{pd} $ in Figs.~\ref{fig:5}(c, d). 
We think these two branches correspond to the harmonic oscillation motion of one single particle and two combined particles within the bottom of the potential well. Note that the oscillation of the two combined particles within the potential well is
similar to the sloshing mode~\cite{Li:2018}.
The oscillation frequencies of a single particle and two combined particles are $\omega_1 =\sqrt {k_s / m}= \sqrt{ 4 \pi^{2} U_0/ m w^2 }$ and $\omega_2 =\sqrt {k_s / 2m}= \sqrt{ 2 \pi^{2} U_0/ m w^2 }$, respectively. 
Thus, under the substrate depth of $U_0=0.5E_0$, the derived frequencies are $0.91 \omega_{pd}$ and $0.64 \omega_{pd}$, respectively. 
Similarly, the substrate depth of $E_0$ results in the two frequencies of $1.28 \omega_{pd}$ and $0.91 \omega_{pd}$, respectively.
Clearly, for each of these two substrate depths, the two derived frequencies agree well with the frequencies in the spectra of Fig.~\ref{fig:5}.

Besides the two branches studied above, there is one more branch in each panel of Fig.~\ref{fig:5}. Since the two branches described above correspond to the harmonic oscillation of a single particle and two combined particles together, the left branch should correspond to the relative motion of the two particles in each potential well, similar to a breathing mode~\cite{Li:2018}.
In their relative motion, these two particles always move
along or perpendicular to the
vector connecting the particles,
i.e., in the radial or the azimuthal directions. Thus, the highest branch of $\tilde{C}_{L}({\bf k}_x,\omega)$ in Fig.~\ref{fig:5}(a) with $\omega = 1.62 \omega_{pd}$ and the lowest branch of $\tilde{C}_{T}({\bf k}_x,\omega)$ in Fig.~\ref{fig:5}(b) with $\omega = 0.42 \omega_{pd}$ should correspond to the relative motion of the two particles in each potential well for the conditions of $U_0 =0.5E_0$ and $\rho = 2$ in the radial and azimuthal directions, respectively. When the substrate depth increases to $U_0 =E_0$, these two frequencies are changed to $2.24 \omega_{pd}$ and $0.50 \omega_{pd}$ as shown in Figs.~\ref{fig:5}(c) and \ref{fig:5}(d), respectively.

\begin{figure}[htb]
    \centering
    \includegraphics{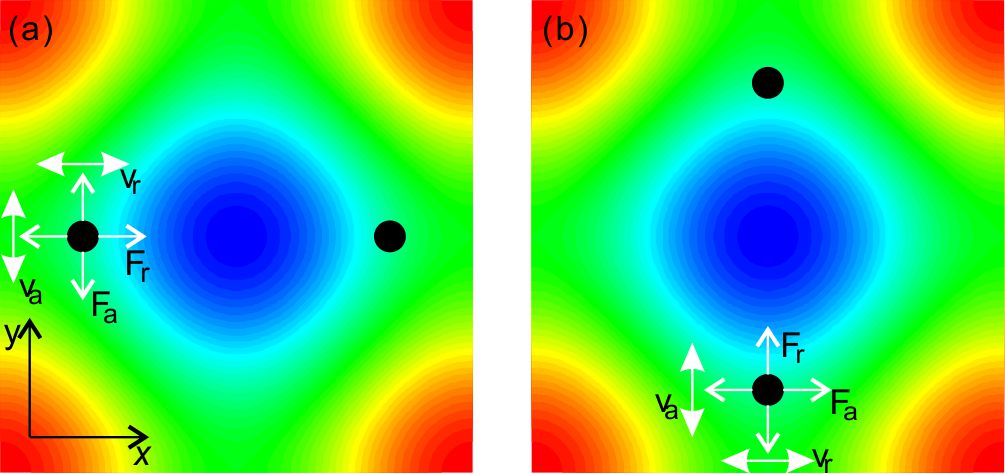}
    \caption{\label{fig:6}
      Schematic of of the two-particle arrangement under the square substrate. The alignment of the two particles within one potential well is in either the $x$ (a) or $y$ (b) directions, i.e.,
parallel to one of the two axes of the square substrate. Each particle oscillates in the radial and azimuthal directions, as $v_r$ and $v_a$ presented here. In the radial direction, the force $F_r$ acting on each particle is the summation of the interparticle repulsion and the restoring force from the substrate. In the azimuthal direction, the restoring force from the substrate is partially canceled out by the interparticle repulsion, due to the purely repulsive force between particles, shown here as $F_a$. Note, to present the forces and velocities clearly, the distance between the two particles is magnified in this schematic.
    }
\end{figure}

We can directly derive these frequencies from the relative motion of the two-body structure inside the potential well in the radial and azimuthal directions, as shown in the schematic of Fig.~\ref{fig:6}. Since these two particles mainly oscillate with small amplitudes around the nearly fixed equilibrium positions, we follow the 1D chain model~\cite{Liu:2005} to linearize the interparticle Yukawa repulsive force to obtain the radial and azimuthal spring constants of $k_r$ and $k_a$ as
\begin{equation}\label{equal_9}
{	k_r = \frac{Q^2( L^2 \kappa^2 +2L \kappa +2)} {4 \pi \epsilon_0 L^3 a^3 e^{L\kappa}} ,}
\end{equation}
\begin{equation}\label{equal_10}
{	k_a = \frac {Q^2(L \kappa +1)}{4 \pi \epsilon_0 L^3 a^3 e^{L\kappa}} ,}
\end{equation}
where $La$ is the distance between these two particles inside the potential well when they are in their equilibrium positions. By incorporating the motion of the particles inside the potential well into their relative motion, using Eqs.~(\ref{equal_9}) and (\ref{equal_10}), we obtain the corresponding oscillation frequencies as
\begin{equation}\label{equal_11}
\begin{aligned}
	\omega_r &= \sqrt{\frac{k_s}{2m} + \frac{k_r}{m/2} } \\
&=\sqrt{ \frac{2 \pi^{2} U_0} {m w^2} + \frac{Q^2(L^2 \kappa^2 +2L \kappa +2)} {2 \pi \epsilon_0 m L^3 a^3 e^{L\kappa}}}
\end{aligned}
\end{equation}
for the radial direction, and
\begin{equation}\label{equal_12}
\begin{aligned}
	\omega_a &= \sqrt{\frac{k_s}{m} - \frac{k_a}{m/2} } \\
&=\sqrt{ \frac{4 \pi^{2} U_0} {m w^2} - \frac{Q^2(L \kappa +1)} {2 \pi \epsilon_0 m L^3 a^3 e^{L\kappa}}}
\end{aligned}
\end{equation}
for the azimuthal direction of the two-body structure in the potential well. In the radial direction, the frequency $\omega_r$ comes from the restoring force of two combined particles from the substrate $\sqrt{k_s/2m}$, coupled with the oscillation from their interparticle repulsion of their reduced mass $\sqrt{k_r/(m/2)}$, since the relative motion of the two particles is in opposite directions. However, in the azimuthal direction, the relative motion refers to the particle motion perpendicular to the radial direction, i.e., the single particle motion behavior, corresponding to
$\sqrt{k_s/m}$, coupled with the
repulsion of their reduced mass $\sqrt{k_a/(m/2)}$. Furthermore, the pure repulsion between particles further increases the force in the radial direction, while decreasing the force in the azimuthal direction, as in Eqs.~(\ref{equal_11}) and (\ref{equal_12}) shown above.

We find that Eqs.~(\ref{equal_11}) and (\ref{equal_12})
correspond to the highest frequency in the longitudinal spectra and the lowest frequency of the transverse spectra. Substituting the substrate depth $U_0=0.5 E_0$ into Eqs.~(\ref{equal_11}) and (\ref{equal_12}), we obtain the derived frequencies of the relative motion as $1.55 \omega_{pd}$ and $0.43 \omega_{pd}$ for the relative motion of the two-body structure in the potential well in the radial and azimuthal directions, which are very close to the two frequencies of $1.62 \omega_{pd}$  and $0.42 \omega_{pd}$ in Figs.~\ref{fig:5}(a) and \ref{fig:5}(b).
When the substrate depth increases to $E_0$, we obtain these two frequencies as $2.20 \omega_{pd}$ and $0.51 \omega_{pd}$, respectively. These two frequencies are also very close to $2.24 \omega_{pd}$ and $0.50 \omega_{pd}$ observed in Figs.~\ref{fig:5}(a) and \ref{fig:5}(b), respectively. Thus, our derived results above agree well with the spectra frequencies in Fig.~\ref{fig:5}. 

Note that, for our
system,
the alignment direction of two particles in either the $x$ or $y$ direction
inside the potential well
greatly simplifies the spectra results in Fig.~\ref{fig:5}.
Two particles aligned in the $y$ direction as shown in Fig.~\ref{fig:6}(b)
mainly oscillate around their equilibrium positions,
so their $x$ coordinates are nearly the same. From Eqs.~(\ref{equal_6}) and (\ref{equal_7}), their relative motion has almost no contribution in the current autocorrelation functions $C_{L}({\bf k}_x,t)$ and $ C_{T}({\bf k}_x,t)$. Thus, only the motion of the pairs of two particles aligned in the $x$ direction
provide substantial contributions to the longitudinal $\tilde{C}_{L}({\bf k}_x,\omega)$ and transverse spectra $\tilde{C}_{T}({\bf k}_x,\omega)$, as shown in Fig.~\ref{fig:6}(a).

\begin{figure}[htb]
    \centering
    \includegraphics{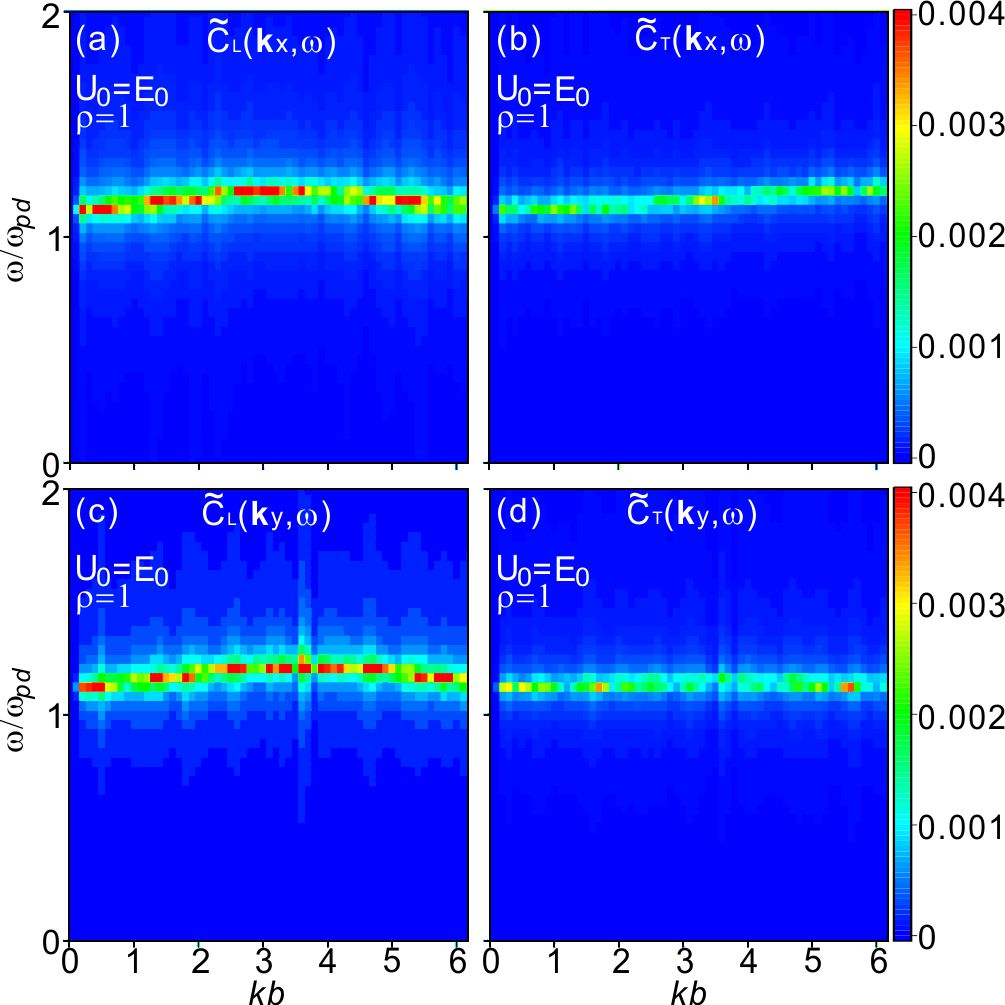}
    \caption{\label{fig:7}
      Calculated longitudinal $\tilde{C}_{L}({\bf k}_x,\omega)$ (a), $\tilde{C}_{L}({\bf k}_y,\omega)$ (c) and transverse phonon spectra $\tilde{C}_{T}({\bf k}_x,\omega)$ (b), $\tilde{C}_{T}({\bf k}_y,\omega)$ (d) of the 2D Yukawa solid under the periodic triangular substrate with $U_0=E_0$ at a commensurability ratio of $\rho = 1$. Clearly, all four spectra are mainly concentrated on a specific frequency value,
in good agreement with the oscillation frequency estimated from one single particle within the potential well. The slight difference between the $x$ and $y$ directions, especially at higher wavenumber values, is mainly due to the anisotropy of the triangular substrate, similar to the hexagonal lattice arrangement of 2D solids. 
    }
\end{figure}

\begin{figure}[htb]
    \centering
    \includegraphics{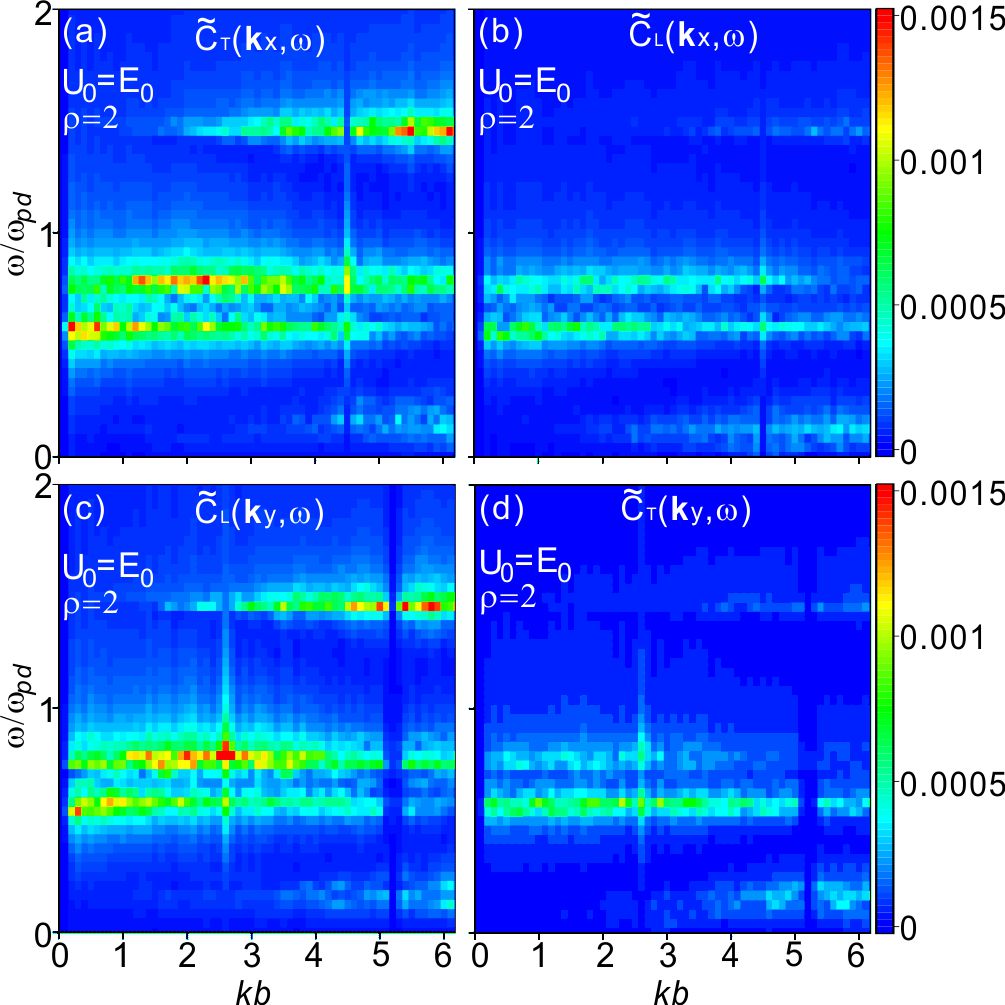}
    \caption{\label{fig:8}
      Calculated longitudinal $\tilde{C}_{L}({\bf k}_x,\omega)$ (a), $\tilde{C}_{L}({\bf k}_y,\omega)$ (c) and transverse phonon spectra $\tilde{C}_{T}({\bf k}_x,\omega)$ (b), $\tilde{C}_{T}({\bf k}_y,\omega)$ (d) of the 2D Yukawa solid under the periodic triangular substrate of $U_0=E_0$, with a commensurability ratio of $\rho = 2$. Clearly, all four spectra are mainly concentrated on
the same four frequencies due to the two-body structure formed by two dust particles in each potential well. The two moderate frequencies agree well with the oscillation frequencies estimated from one single particle and two combined particles within the potential well. The highest and lowest frequencies are in good agreement with the oscillation frequencies of the relative motion of the two-body structure in the potential well, in the radial and azimuthal directions, respectively. 
    }
\end{figure}

\subsection{{\bf C}. Wave spectra under a triangular substrate}
We next study the effect of a triangular substrate on the phonon spectra of the 2D Yukawa solid.
Figure~\ref{fig:7} presents the phonon spectra of the 2D Yukawa solid under the triangular substrate with a depth of $U_0 = E_0$ at a commensurability ratio of $\rho = 1$.
Clearly, for all four spectra in Fig.~\ref{fig:7}, the nearly zero slope indicates that the group velocity is nearly zero, i.e., the wave propagation is strongly suppressed.
Although the current 2D Yukawa solid
is strongly anisotropic in the $x$ and $y$ directions
as shown in Fig.~\ref{fig:2}(c), the calculated longitudinal spectra $\tilde{C}_{L}({\bf k}_x,\omega)$ in Fig.~\ref{fig:7}(a) are nearly the same as $\tilde{C}_{L}({\bf k}_y,\omega)$ in Fig.~\ref{fig:7}(c) for almost
the entire range of wavenumbers.
However, only
for small wavenumbers,
the transverse spectra $\tilde{C}_{T}({\bf k}_x,\omega)$ in Fig.~\ref{fig:7}(b) are the same as $\tilde{C}_{T}({\bf k}_y,\omega)$ in Fig.~\ref{fig:7}(d). In the higher wavenumber range, the transverse phonon spectra are slightly different, i.e., the frequency of $\tilde{C}_{T}({\bf k}_x,\omega)$ increases slightly with the wavenumber, while
the frequency of $\tilde{C}_{T}({\bf k}_y,\omega)$ does not,
due to the anisotropy of our
system, as shown in Fig.~\ref{fig:2}(c). 

Interestingly, we find that all four spectra in Fig.~\ref{fig:7} have only one frequency, which is $\approx 1.12 \omega_{pd}$, suggesting that all particles mainly oscillate in both directions with this frequency inside the potential well.
This frequency can also be derived from the harmonic oscillation motion of a single particle within the triangular potential well of Eq.~(\ref{equal_3}).  
We can linearize the force from the triangular potential well, Eq.~(\ref{equal_5}), to obtain the corresponding spring constant of $k_t =  16 \pi^{2} U_0/ 9 w^2 $, where the subscript $t$ refers to the triangular potential well. 
Thus, we obtain the oscillation frequency of a single particle as $\omega_1 =\sqrt {k_t / m}= \sqrt{ 16 \pi^{2} U_0/ 9 m w^2 }$. 
Substituting the depth $U_0$ and width $w$ of the potential well, we derive the oscillation frequency of $1.13 \omega_{pd} $ for the depth of $E_0$,
in good agreement with the phonon spectra frequency in Fig.~\ref{fig:6}.

In Fig.~\ref{fig:8}, we present the phonon spectra of the 2D Yukawa solid under the triangular substrate with the same substrate depth $E_0$, where the commensurability ratio is changed to $\rho = 2$. Clearly, the longitudinal and transverse phonon spectra in Fig.~\ref{fig:8} contain four branches, similar to the spectra in Fig.~\ref{fig:5}. The frequency of each branch is almost unchanged while the wavenumber varies, also indicating that the corresponding wave propagation is strongly suppressed in Fig.~\ref{fig:8}. 
In the four panels of Fig.~\ref{fig:8}, the frequency values of the four branches are almost the same, which are $1.45 \omega_{pd}$, $0.79 \omega_{pd}$, $0.58 \omega_{pd}$ and $0.17 \omega_{pd}$, respectively.
Similar to Fig.~\ref{fig:5}, the two moderate frequencies in Fig.~\ref{fig:8} probably come from the oscillation of one single particle and two combined particles inside the potential well of the triangular substrate.
Using the spring constant of the triangular substrate with the depth of $U_0=E_0$ obtained above, the oscillation frequencies of a single particle and two combined particles are $\omega_1 =\sqrt {k_t / m}= \sqrt{ 16 \pi^{2} U_0/ 9 m w^2 } = 0.81 \omega_{pd}$ and $\omega_2 =\sqrt {k_t / 2m}= \sqrt{ 8 \pi^{2} U_0/ 9 m w^2 } = 0.57 \omega_{pd}$, respectively. 
Clearly, these two derived frequency values agree well with the two moderate phonon spectra frequencies in Fig.~\ref{fig:8}. Note that the oscillation of the two combined particles inside one potential well
corresponds to the sloshing mode~\cite{Li:2018}.

Besides the two moderate frequency values above, the highest and lowest frequencies in Fig.~\ref{fig:8} probably come from the relative motion of the two-body structure in the triangular potential well, in the radial and azimuthal directions, respectively. Due to the alignment
of the two particles in each triangular potential well
along  $\approx 60^{\circ}$ or $120^{\circ}$ with respect to the $x$ direction, their relative motion makes contributions to both the longitudinal and transverse spectra. After incorporating the derived spring constants $k_r$ and $k_a$ of the relative motion in the radial and azimuthal direction from the interparticle Yukawa repulsion, Eqs.~(\ref{equal_9}) and (\ref{equal_10}), into the triangular potential well, we can derive the oscillation frequencies as
\begin{equation}\label{equal_13}
\begin{aligned}
	\omega_r &= \sqrt{\frac{k_t}{2m} + \frac{k_r}{m/2} } \\
&=\sqrt{ \frac{8 \pi^{2} U_0} {9 m w^2} + \frac{Q^2(L^2 \kappa^2 +2L \kappa +2)} {2 \pi \epsilon_0 m L^3 a^3 e^{L\kappa}}}
\end{aligned}
\end{equation}
for the radial direction, and
\begin{equation}\label{equal_14}
\begin{aligned}
	\omega_a &= \sqrt{\frac{k_t}{m} - \frac{k_a}{m/2}} \\
&=\sqrt{ \frac{16 \pi^{2} U_0} {9 m w^2} - \frac{Q^2(L \kappa +1)} {2 \pi \epsilon_0 m L^3 a^3 e^{L\kappa}}}
\end{aligned}
\end{equation}
for the azimuthal direction of the two-body structure in the potential well. Clearly, in Eqs.~(\ref{equal_13}) and (\ref{equal_14}), the coupling with the relative motion of the two-body structure from the interparticle repulsion of their reduced mass is exactly the same as $\sqrt{k_r/(m/2)}$ in Eq.~(\ref{equal_11}) and $\sqrt{k_a/(m/2)}$ in Eq.~(\ref{equal_12}). Substituting the depth $U_0=E_0$ into Eqs.~(\ref{equal_13}) and (\ref{equal_14}), we obtain the oscillation frequencies of $1.51 \omega_{pd}$ and $0.18 \omega_{pd}$ for the relative motion of the two-body structure in the potential well, in the radial and azimuthal directions, respectively. Clearly, these two derived frequencies agree well with the phonon spectra frequencies shown in Fig.~\ref{fig:8}. Note that the spectra with the higher frequency of $1.51 \omega_{pd}$
corresponds to the breathing mode~\cite{Li:2018}.

\section{\uppercase\expandafter{\romannumeral4}. Summary}
We investigate the phonon spectra of a 2D solid dusty plasma modified by 2D square and triangular periodic substrates using Langevin dynamical simulations. We find that the wave propagation is strongly suppressed due to the confinement of the particles by the applied 2D substrates.
When the commensurability ratio is $\rho = 1$, i,e, only one particle inside each potential well, the spectra mainly concentrate on one specific frequency for all studied wavenumbers, agreeing well with our derived harmonic frequency of one single particle oscillation inside one potential well. 
When the commensurability ratio is $\rho=2$, corresponding to two particles on average within each potential well, the
longitudinal and transverse spectra
split into four branches in total, where the frequency value for each branch is nearly unchanged for all wavenumbers. The two moderate frequencies can be derived from the harmonic oscillation frequency values of one single particle and two combined particles inside the potential well, respectively. The frequencies of the other two branches can be derived from the relative motion of the two-body structure inside one potential well, in the radial and azimuthal directions, respectively. The force amplitude from the potential well and the interparticle Yukawa repulsion both determine these frequency values. The difference between the spectra results modified by the square and triangular substrates comes from the anisotropy of substrates and the resulting alignment directions of the two-body structure inside the potential wells. There are several future directions to examine, including the ordering and band gaps that appear at high fillings such as three or four particles per trap, fractional fillings, and examining other substrate symmetries or even a quasiperiodic substrate to determine how the band gaps change. In fact, there have already been several works examining colloidal ordering on quasiperiodic substrates, for example in Ref.~\cite{mikhael:2008}.

\section{Acknowledgments}

Work was supported by the National Natural Science Foundation of China under Grant No. 11875199, the 1000 Youth Talents Plan, startup funds from Soochow University, the Priority Academic Program Development (PAPD) of Jiangsu Higher Education Institutions, and the US Department of Energy through the Los Alamos National Laboratory. Los Alamos National Laboratory is operated by Triad National Security, LLC, for the National Nuclear Security Administration of the U. S. Department of Energy (Contract No. 892333218NCA000001).

\end{document}